\documentclass[doublecol]{epl2} 
\usepackage{graphicx,amsmath,amssymb,bm,color,engord}
\usepackage[ocgcolorlinks,colorlinks,urlcolor=blue,citecolor=blue,linkcolor=blue]{hyperref}

\title{Transfer and  storage of  qubits  in the presence of decoherence}
\author{Kelly R. Patton and Uwe R. Fischer}
\institute{Seoul National University, Department of Physics and Astronomy\\ Center for Theoretical Physics, 151-747 Seoul, Korea }


\abstract{The effects of decoherence on the transfer and storage of coherent quantum states  
in hybrid systems are studied within the Caldeira-Leggett approach.  In general, we find that a 
high transfer fidelity can be achieved even if the decoherence time is less than an order of magnitude larger than the transfer time, which is approximately half a Rabi period and determined by the qubit-qubit coupling strength.  Finally, we apply our  results to assess the feasibility of a hybrid quantum memory system, comprised of the hyperfine qubit states of an ultracold atomic Bose-Einstein condensate and the flux qubit of a SQUID.}

\begin{document}

\maketitle


The implementation  of a quantum analog of random-access memory for qubits, qRAM, is currently one of the major goals of quantum information processing \cite{Blencowe}.  As with their classical counterpart, which can be found in all modern-day computers, the qRAM should be both fast and robust. Ideally, arbitrary  coherent quantum states can be quickly passed to, stored, and retrieved from the qRAM with high fidelity.  Unlike  in classical systems, these processes  have to be performed with an unavoidable continuous loss of information, caused by 
the detrimental influence of decoherence. The minimization of the effects of decoherence on transfer and storage of quantum information is therefore at the forefront of implementing a qRAM architecture and of quantum computing in general. To realize this demanding goal, quantum-hybrid systems represent some of the most interesting and promising candidates \cite{Blais,Verdu,Lukin,Armour,Nakamura,Wallraff}. 

Within a hybrid architecture one subsystem, or qubit, works as part of the quantum central processing unit (CPU), while the other component operates as the qRAM.  One of the major benefits of combining  seemingly disparate systems is that they can each reinforce various advantages, while compensating for deficiencies, of either subsystem.  For example, recent experiments involving the nuclear and  electron spin states of nitrogen-vacancy (NV) centers in diamond, and a sophisticated dynamical decoupling of qubits,  have produced record decoherence times for a solid-state system \cite{vanderSar}, on the order of a second.   Along with constructing a qRAM architecture from either a single NV center \cite{Fuchs}  or an ensemble of such centers \cite{Zhu,Wu}, a  variety of other hybrid systems have  been experimentally realized, such as superconducting  qubits or atomic/molecular quantum gases coupled to electromagnetic and nanomechanical resonators \cite{Majer, OConnell, Cleland, Rabl,Brennecke}.

By their very nature, each component of a hybrid system can have vastly different coherence times.   Frequently, as is the case in many of the previous examples, the CPU  element originates from a solid-state system, where the qubit states, $|0\rangle$ and $|1\rangle$, span the low-energy  Hilbert space of some effective quasiparticle.  While solid-state qubits have many advantages, such as scalability  and controllability, they typically have small coherence times. This is  a major hurdle for quantum information processing.  In contrast, the memory qubits may not be as easily constructed on a large scale,  but they potentially have much longer decoherence times, making them ideal for storage of quantum information.

The transfer of a quantum state from one qubit to the other is commonly done in experimentally realized hybrids via the Rabi process, cf. e.g. \cite{vanderSar,OConnell}, dynamically controlling one qubit's level spacing, bringing the two subsystems into and out of resonance for half a Rabi cycle.  The Rabi period is determined by the qubit-qubit interaction.  Ideally this coupling should be strong enough to enable a fast transfer,  but weak enough to not excite states outside of the qubit Hilbert space.
Although there has been previous work on the decay of Rabi oscillations between two qubits caused by decoherence \cite{AmentPRB06} and on qubit state transfer in the presence of decoherence in the perturbative regime \cite{EscherJPhysB11},  a general study on the effects of decoherence during a complete qRAM memory operation, i.e., state transfer and storage, is however missing. In what follows, we undertake such a study, determining a practically applicable lower bound of the transfer fidelity which can be obtained as a function of the ratio of decoherence to transfer time.
This bound is of fundamental importance for the development of current and future hybrids, when the latter ratio becomes of order unity. 

We simulate the transfer and storage of a state from one qubit to another in the  presence of decoherence and determine the transfer fidelity.  Within the Caldeira-Leggett model of dissipation and generalized spin-boson coupling \cite{Leggett} considered here, we find the limiting decoherence time is determined by the energy-relaxation time $T_{1}$.  Naturally, 
one would expect that to obtain a high fidelity transfer, the transfer time should be much smaller than the coherence time. However, we find that to obtain a fidelity  greater than 95\%,  a $T_{1}$ time of  only about three times larger than the qubit-qubit coupling 
Rabi period, which in turn is larger than the transfer time by a factor between two and five (see also below), is required.   As a direct application of these results to a concrete system, we propose a hybrid qRAM system, composed of a flux qubit (SQUID) and the hyperfine states of a  trapped atomic Bose-Einstein condensate (BEC). With current technology, the relevant time scales of decoherence and Rabi period in this hybrid can be of the same order.  Therefore, the inclusion of decoherence effects in a state transfer simulation are needed to verify the feasibility of this qRAM architecture.  

For the coupled qubit-qubit system, the initial state is prepared in a parent qubit (p) and passed to a memory  qubit (m).  The  Hamiltonian is taken to be $(\hbar=1)$
\begin{equation}
\label{qubit-qubit Hamiltonian}
 \hat{H}_{\rm q\text{-}q}(t)=\frac{\omega_{\rm m}}{2}\sigma^{}_{z}\oplus\frac{\omega_{\rm p}(t)}{2}\sigma^{}_{z}-\left(\begin{array}{cc}0 & \Omega \\\Omega^* & 0\end{array}\right)\otimes \sigma^{}_{x},
\end{equation}
where  $\hat{A}\oplus\hat{B}=\hat{A}\otimes \mathbb{I}+\mathbb{I}\otimes\hat{B}$ with $\mathbb{I}$ being the identity matrix, $\sigma_{i}$ are  Pauli matrices, and the time dependence of  the level spacing of the parent qubit, $\omega_{\rm p}(t)=\omega_{\rm m}W(t)$, is used to bring the two qubits into and out of resonance \cite{windowfunction}.  The Rabi coupling $\Omega$ is assumed complex, due to the symmetrical $\sigma_{x}\otimes\sigma_{x}$ and $\sigma_{y}\otimes\sigma_{x}$ interaction terms. Although the form of the qubit-qubit coupling is chosen to coincide with the hybrid system proposed below, the following method and results can be easily  generalized to include other forms. 

 There can be many sources of decoherence, originating both internally and externally.  These sources will be  collectively  modeled phenomenologically, by the common Caldeira-Leggett type 
 method of coupling the qubit's degrees of freedom to a bosonic bath \cite{Leggett}.  Assuming  the decoherence of the parent qubit is the dominant limiting factor on state transfer, such that the decoherence of the memory qubit can be neglected in comparison, we only consider the parent qubit to be directly coupled to a bath.  The qubit-bath coupling is taken to be
\begin{equation}
\label{qubit-bath coupling}
\hat{V}=\sum_{\bm k}\boldsymbol{\lambda}^{}_{\bm k}\cdot\boldsymbol{\sigma}\otimes\big(\hat{b}^{}_{\bm k}+\hat{b}^{\dagger}_{\bm k}\big),
\end{equation} 
where $\boldsymbol{\sigma}=(\sigma_{x},\sigma_{y},\sigma_{z})$ and $\boldsymbol{\lambda}^{}_{\bm k}=(\lambda^{x}_{\bm k},\lambda^{y}_{\bm k},\lambda^{z}_{\bm k})$ is the coupling constant to each bath mode, labeled by $\bm k$. 
 The total  Hamiltonian  of the qubit-qubit-bath system is then  
\begin{equation}
\label{Total qubit-qubit-bath Hamiltonian}
\hat{H}_{}(t)=\hat{H}_{\rm q\text{-}\rm q}(t)\oplus\sum_{\bm k}\epsilon^{}_{\bm k}\hat{ b}^{\dagger}_{\bm k}\hat{b}^{}_{\bm k}+\mathbb{I}\otimes \hat{V},
\end{equation}
where $\epsilon_{\bm k}$ determines the dispersion of the bath bosons.

To accurately determine the fidelity of a state transferred in the presence of decoherence, we numerically solve the time-dependent master equation for the reduced density matrix of the two-qubit system $\rho(t)$.  If at time $t=0$ the density matrix of the qubit-qubit-bath system is given by $\tilde{\rho}(0)=\rho_{}(0)\otimes {\rm e}^{-\beta \hat{H_{\rm b}}}/Z_{\rm b}$, with $\beta=(k_{\rm B}T)^{-1}$ (the inverse temperature), $\hat{H}_{\rm b}=\sum_{\bm k}\epsilon^{}_{\bm k}\hat{ b}^{\dagger}_{\bm k}\hat{b}^{}_{\bm k}$, and $Z_{\rm b}={\rm Tr}\, {\rm e}^{-\beta\hat{H}_{\rm b} }$, then the full density matrix at a later time is given by $\tilde{\rho}(t)=\hat{U}_{H}(t)\tilde{\rho}(0)\hat{U}^{\dagger}_{H}(t)$, where the time-evolution operator $\hat{U}_{H}(t)$ is the time-ordered exponential of the full Hamiltonian \eqref{Total qubit-qubit-bath Hamiltonian}.  Changing to the interaction representation, $\tilde{\rho}_{\rm I}(t)=\hat{U}^{\dagger}_{H_{0}}(t)\tilde{\rho}(t)\hat{U}_{H_{0}}(t)$, where $\hat{H}_{0}(t)=\hat{H}_{\rm q\text{-}\rm q}(t)\oplus \hat{H}_{\rm b}$ and $\hat{U}_{H_{0}}(t)=\hat{U}_{H_{\rm q\text{-}\rm q}}(t){\rm e}^{-i\hat{H}_{\rm b}t}$.  Within the Markov approximation \cite{BreuerBook} the equation of motion can  be written 
\begin{equation}
\label{equation of motion for density matrix}
 \partial_{t}\tilde{\rho}_{\rm I}(t)=-i\big[\hat{V}_{\rm I}(t),\tilde{\rho}_{\rm I}(0)\big]-\int_{0}^{t}dt'\, \big[\hat{V}_{\rm I}(t),\big[\hat{V}_{\rm I}(t'),\tilde{\rho}_{\rm I}(t)\big]\big]. 
\end{equation}
In addition, we also make the Born approximation, which neglects correlations  that develop between the qubit subsystem and the bath. In other words, at later times the density matrix is taken to be of the form $\tilde{\rho}_{}(t)\simeq\rho(t)\otimes {\rm e}^{-\beta \hat{H}_{\rm b}}/Z_{\rm b}$.   An equation of motion for the reduced density matrix can then be found from \eqref{equation of motion for density matrix} by tracing over the bath; $\rho_{\rm I}(t)={\rm Tr}_{\rm b} \tilde{\rho}_{\rm I}(t)$.  Assuming an isotropic qubit-bath coupling  $\boldsymbol{\lambda}_{\bm k}=\lambda_{\bm k}(\bm{e}_{x}+\bm{e}_{y}+\bm{e}_{z})\equiv\lambda_{\bm k}\bm{n}$  and ${\rm Tr}_{\rm  b}{\rm e}^{-\beta \hat{H}_{\rm b}}\hat{b}^{(\dagger)}_{\bm k}(t)=0$, the equation of motion for the reduced density matrix is given by  the equation 
\begin{align}
\label{equation for the reduced density matrix}
&\partial_{t}\rho^{}_{\rm I}(t)=-\int_{0}^{t}dt'\int_{0}^{\infty}d\epsilon\, J(\epsilon)\Big\{\coth(\epsilon\beta/2)\cos(\epsilon(t-t'))\nonumber\\&\times \big[\mathbb{I}\otimes\bm{n}\cdot \boldsymbol{\sigma}_{\rm I}(t),\big[\mathbb{I}\otimes\bm{n}\cdot \boldsymbol{\sigma}_{\rm I}(t'),\rho^{}_{\rm I}(t)\big]\big]-i\sin(\epsilon(t-t'))\nonumber\\&\times \big[\mathbb{I}\otimes\bm{n}\cdot \boldsymbol{\sigma}_{\rm I}(t),\big\{\mathbb{I}\otimes\bm{n}\cdot \boldsymbol{\sigma}_{\rm I}(t'),\rho^{}_{\rm I}(t)\big\}\big]\Big\},
\end{align}
where the time dependence  of the operators is now given by $\hat{U}_{H_{\rm q\text{-}\rm q}}(t)$, and  $J(\epsilon)=\sum_{\bm k}\lambda^{2}_{\bm k}\delta(\epsilon-\epsilon_{\bm k})$ is a coupling weighted density of states or spectral density of the bath.  A common choice of $J(\epsilon)$ is  the so-called ohmic bath, with $J(\epsilon)\propto \epsilon$  up to some upper  cutoff $\omega_{\rm c}$.  The form used here is $J(\epsilon)=\eta \epsilon e^{-\epsilon/\omega_{\rm c}}$, where $\eta$ is the dimensionless dissipation or friction constant.  The energy  integral over $\epsilon$ can be done analytically, but is not important for the following results.  

\begin{figure}[h]
\includegraphics[width=\columnwidth]{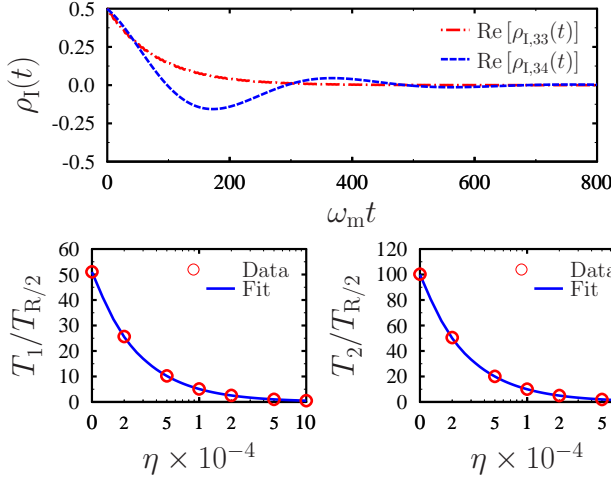}%
\caption{The calculated decoherence times of the parent qubit are shown in units of half the Rabi period $T_{\rm R/2}\equiv\frac{\pi}{2|\Omega|}$, with $|\Omega|=0.01\,\omega_{\rm m}$, as a function of the qubit-bath coupling. At time $t=0$ the system is placed into the state $|\Psi\rangle=2^{-1/2}(|00\rangle+|01\rangle)$ and then time evolved according to Eq.\,\eqref{equation for the reduced density matrix}.  The fit to the $T_{1}$ data is  given by $f(x)=\frac{5.10\times 10^{-4}}{x}$ and for $T_{2}$ by $f(x)=\frac{10^{-3}}{x}$.  The top panel shows an example of the decay of the elements of the density {matrix, $\rho_{\rm I,33}(t)=\langle10|\rho_{\rm I}(t)|01\rangle$ and $\rho_{\rm I,34}(t)=\langle10|\rho_{\rm I}(t)|00\rangle$,} as a function of time. $T_{1}$ is found by fitting the decay to a function of the form $c_{0}e^{-t/T_{1}}$, while $T_{2}$ is extracted by fitting the off-diagonal decay  to $c_{0}\cos(c_{1} t)e^{-t/T_{2}}$.   
 \label{fig1}}
\end{figure}
\begin{figure}[h]
\includegraphics[width=\columnwidth]{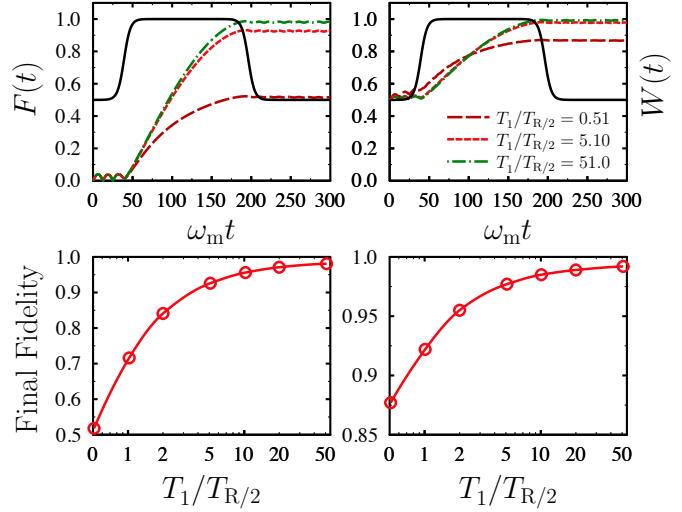}%
\caption{(color online) The left column shows the time-dependent fidelities $F(t)$ (top) and final fidelity (bottom), for the direct transfer $|01\rangle\to |10\rangle$, and various  $T_{1}$ times corresponding to the data in Fig.\,\ref{fig1};  $T_{\rm R/2}$ is half a Rabi period, the approximate time required to transfer a state. The solid line in the top left and right plots is the time profile of $W(t)$ \cite{windowfunction}, the function used to bring the two qubits into and out of resonance; the solid line on the bottom plots is a guide to the eye. To obtain a fidelity of more than 90\% for this direct state transfer requires a $T_{1}\gtrsim 5\, T_{\rm R/2}$. The right column shows the fidelities 
for the transfer of a superposition state $2^{-1/2}\big(|00\rangle+i|01\rangle\big)\to 2^{-1/2}\big(|00\rangle+i|10\rangle\big)$, for the same $T_{1}$ times as in the left column. Similarly to the direct state transfer, to obtain a fidelity of more than 90\% only requires a $T_{1}\gtrsim T_{\rm R/2}$.}\label{fig2}
\end{figure}

Next, we numerically solve the 16 coupled differential equations of \eqref{equation for the reduced density matrix} for the reduced density matrix elements using a fourth-order Runge-Kutta method with a time step of $\omega_{\rm m}t=1$ and  parameters: $\omega_{\rm c}=100\,\omega_{\rm m}$, $\beta=100\,\omega^{-1}_{\rm m}$, $|\Omega|=0.01\,\omega_{\rm m}$.  {The temperature $\beta^{-1}$ is chosen to be much smaller than any qubit energy spacing. This is, first of all, necessary to ensure a well-defined qubit. 
Furthermore, in this regime the dephasing due to thermal effects is negligibly small; as long as $\beta^{-1} \ll \omega_{\rm m}$, our
results will therefore not significantly depend on the value of $\beta$ \cite{tempnote}.} Additionally, the nontrivial time-evolution operator $\hat{U}_{H_{\rm q\text{-}\rm q}}(t)$, which determines the time dependence of the operators in \eqref{equation for the reduced density matrix}, and the final time integral have to be numerically found at each  Runge-Kutta time step.  The Rabi frequency  used for the simulations allows for the calculations to be done in a reasonable time frame, while still within the weak coupling regime. Furthermore,  the physically relevant quantity is the dimensionless  ratio of the  decoherence time to the time needed to transfer a state, approximately half a Rabi period $T_{\rm R/2}\equiv\frac{\pi }{2|\Omega|}$.

Figure \ref{fig1} shows the obtained decoherence times of the parent qubit as a function of the ohmic friction constant $\eta$. These were found by first decoupling the parent qubit from the memory one,  by setting $\Omega=0$ in  \eqref{qubit-qubit Hamiltonian}.  At time $t=0$ the system is then put in the pure state $|\Psi\rangle=2^{-1/2}(|00\rangle+|01\rangle)$, where the computational basis states $|ij\rangle=|i\rangle_{\rm m}\otimes |j\rangle_{\rm p}$ are the energy eigenstates of the uncoupled system.  The energy-relaxation time $T_{1}$ and phase-decoherence time $T_{2}$ are then found by fitting the decay of the diagonal $(T_{1})$ and off-diagonal $(T_{2})$ components of $\rho^{}_{\rm I}(t)$ to a decaying exponential of the form $e^{-t/T_{i}}$ \cite{T1andT2times}. As can be seen from Fig.~\ref{fig1}, within this model $T_{2}\approx 2\,T_{1}$; therefore, the $T_{1}$ energy relaxation time is the major  limiting factor of any state transfer.  The $T_{1}$ time can also be  found by calculating the lifetime of the excited state $\tau$ from diagrammatic many-body perturbation theory. The lifetime is then defined as $\tau^{-1}=-2{\rm Im}\, \Sigma(\omega_{\rm p})$ \cite{Mahan}, where $\Sigma(\omega)$ is the parent qubit's self-energy.  To second order in the qubit-bath coupling, which contains the leading  term describing the relaxation of the qubit  and excitation of a boson in the bath,   $T_{1}\simeq\tau=(4\pi \eta \omega_{\rm p})^{-1}e^{\omega_{\rm p}/\omega_{\rm c}}$, which agrees well with the numerically calculated values from our model, as  
shown in Fig.\,\ref{fig1}. 

To transfer a quantum state from the parent qubit to the other, one first initializes the system by bringing the two qubits' level spacings out of resonance. Then, one prepares the parent qubit in an arbitrary initial  state $|\Psi_{\rm i}\rangle=\alpha|00\rangle+\beta|01\rangle$.  Next, the two subsystems are brought into resonance for half a Rabi period $T_{\rm R/2}$. Finally, one takes the two qubits out of resonance, leaving the parent qubit in its ground state and its original state in the memory qubit $|\Psi_{\rm f}\rangle=\alpha|00\rangle+\beta|10\rangle$. The {\em transfer fidelity} is defined to be $F(t)=\sqrt{\langle \Psi_{\rm f}(t)|\rho(t)|\Psi_{\rm f}(t)\rangle}$, where $\rho(t)$ is the reduced density matrix of the qubit system transformed back to the Heisenberg picture, and $|\Psi_{\rm f}(t)\rangle$ is the final target state,  time evolved under the time dependent but decoupled qubit-qubit Hamiltonian $H_{\rm q\text{-}\rm q}$, Eq.\,\eqref{qubit-qubit Hamiltonian}.

Figure \ref{fig2} shows the obtained time evolution of the fidelity and the final fidelities immediately 
after the transfer has completed, for various $T_{1}$ times, and for a direct $|01\rangle\to |10\rangle$ (left column) 
and  superposition  state (right column) transfer.  As one can see, the fidelity is strongly dependent on the initial state. In particular to transfer the superposition state $2^{-1/2}\big(|00\rangle+i|01\rangle\big)\to 2^{-1/2}\big(|00\rangle+i|10\rangle\big)$ a fidelity greater than 90\% can readily be achieved for $T_{1}\simeq T_{\rm R/2}$; while for the transfer $|01\rangle\to |10\rangle$ a fidelity of only about 50\% is found.  This is primarily due to the fact that for the superposition state the initial wave function has already a significant overlap with the final transferred one. Thus, even before any transfer steps are taken, it starts with a 50\% fidelity.    Generally, we find that to obtain a fidelity of at least  95\%, $T_{1}$ should be 
no smaller than about three times the Rabi period. 
Additionally,  because of the bath, after the transfer the state can still slowly dissipate from the memory qubit---even when the two subsystems are out of resonance.  This occurs on a time scale on the order of $(\omega_{\rm m}/|\Omega|)^{2}\,T_{1}$. This finite lifetime comes from a process where the memory qubit state $|1\rangle$ can relax, exciting the parent qubit's state from its ground state, which then decays into the bath.  In diagrammatic many-body theory, this comes from a fourth-order term in the expansion  of the memory qubit's self-energy. {We conclude that in general the decoherence, due to the bath in which the parent system 
is immersed, must be taken into account also {\it after} the qubit has been transferred to the qRAM. However, in certain physical realizations, see, e.g., Ref.~\cite{vanderSar}, parent and memory qubits can be dynamically decoupled, $\Omega\rightarrow 0$, 
eliminating such decoherence effects.}

The immediate applicability of our results can be seen in the following quantum-hybrid qRAM architecture.  The memory qubits consist of the hyperfine split ground state of a trapped ultracold atomic BEC, e.g, $^{87}{\rm Rb}$.  These states have significant potential for a qRAM, as their $T_{1}$ times are essentially infinite, on the order at least of seconds, and potentially even much longer than the lifetime of the BEC itself.  Recent experimental data for a BEC cloud trapped near a superconducting waveguide resonator geometry indeed indicates phase-coherence times $T_2\sim 7$ sec \cite{Bernon},  
so that the use of our presently proposed hybrid for long-term storage of quantum information is promising \cite{Note2}.

\begin{figure}
\includegraphics[width=\columnwidth]{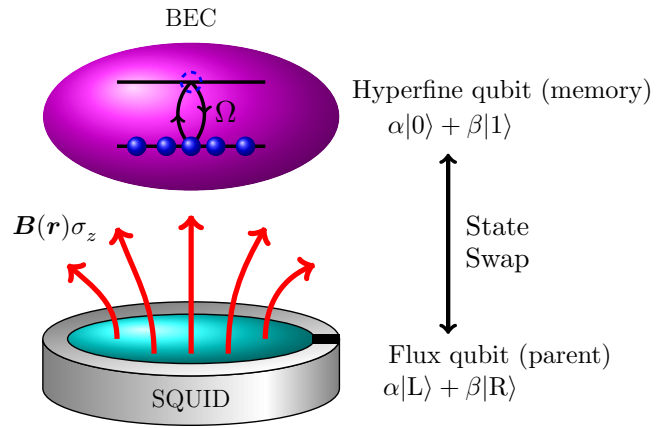}%
\caption{\label{fig4} A schematic representation of a SQUID-BEC hybrid {(not to scale)}. The flux qubit is created from superposing left- and right-going current states of the SQUID,  coupled by its magnetic field to the hyperfine transition of an atomic BEC, where the Rabi process periodically flips one atom from a given hyperfine spin state to the other.}
\end{figure}

The BEC states are magnetically coupled to the parent flux qubit of a SQUID \cite{Makhlin}, which during a Rabi cycle excites a single atom from the macroscopically occupied ground state containing $N$ atoms ($\downarrow$) to an unoccupied excited state ($\uparrow$), see Fig.~\ref{fig4}.
The qubit subspace of the BEC is spanned by
\begin{align}
\label{BEC basises}
|0\rangle^{}_{\rm B}=&(N!)^{-1/2}\big(\hat a^{\dagger}_{\downarrow}\big)^{N}|{\rm vac}\rangle, \nonumber\\
|1\rangle^{}_{\rm B}=&[(N-1)!]^{-1/2}\big(\hat a^{\dagger}_{\downarrow}\big)^{N-1}\hat a^{\dagger}_{\uparrow}|{\rm vac}\rangle.
\end{align}
Using the left-right current states of the SQUID \cite{Makhlin}, the uncoupled BEC-SQUID Hamiltonian then reads
\begin{align}
\label{uncoupled BEC-SQUID Hamiltonian}
\hat H_{\rm BS}=\frac{\omega_{\rm hf}}{2}\sigma^{}_{z}\oplus\left(\frac{\varepsilon}{2}\sigma^{}_{z}-\frac{\Delta(t)}{2}\sigma^{}_{x}\right),
\end{align}
where $\omega_{\rm hf}$ ($=\omega_{\rm m}$ in Eq.\,\eqref{qubit-qubit Hamiltonian}) is the hyperfine splitting of the BEC states (for $^{87}{\rm Rb}$, $\omega_{\rm hf}\approx 6.8\,$GHz), and $\varepsilon$ is the energy difference of the $|\rm L\rangle$ and $|\rm R\rangle$ current states with tunneling amplitude $\Delta(t)$, which can be dynamically modulated to bring the two systems into and out of resonance \cite{Paauw}. 
{In Eq.\,\eqref{uncoupled BEC-SQUID Hamiltonian}, the effective qubit-BEC Hamiltonian, 
$\hat{H}_{\rm B}=\frac{\omega_{\rm hf}}{2}\sigma_{z}$, is expressed in the qubit (i.e. energy eigenstate) basis \eqref{BEC basises}, while the SQUID Hamiltonian is in the more commonly used $|\rm L\rangle$, $|\rm R\rangle$ basis. On the other hand, for $\varepsilon=0$, in the qubit basis  the SQUID Hamiltonian is simply given by $\hat{H}_{\rm S}=\frac{\Delta(t)}{2}\sigma_{z}$, and $\omega_{\rm p}=\Delta(t)$ in \eqref{qubit-qubit Hamiltonian}, then.}

In the left-right current state basis of the SQUID, a  current operator can be defined as follows;  $\hat{I}\simeq I\sigma_{z}$, where $I=\langle{\rm L}|\hat I(\hat\Phi)|{\rm L}\rangle\simeq -\langle{\rm R}|\hat I(\hat \Phi)|{\rm R}\rangle$\ (exact equality holding when $\epsilon =0$). The current operator is related to the flux operator  by $\hat I =(\hat \Phi-\Phi^{}_{\rm ex})/{L}$ with $ \Phi^{}_{\rm ex}$ the classical externally applied flux and $L$ the loop inductance.  The magnetic vector potential of a loop of radius $R$ 
in the dipole approximation and  polar coordinates reads  $\hat {\bm A} \simeq - \mu^{}_{0}R^{2} I\sin\theta/(4r^{2})\,  {\bm e}_{\phi} \,\sigma_{z}
$ \cite{operatornote}.   The  SQUID's magnetic field operator is then $\hat {\bm B} = \nabla \times \hat {\bm A}= {\bm B}({\bm r}) \sigma_{z}$.

For an atom with total magnetic moment $\bm \mu$, (including  orbital, nuclear, and electronic spin contributions) the magnetic dipole coupling of the BEC and SQUID is  given by $H_{\rm int}=-\int d{\bm r}\, \hat{{\bm m}}({\bm r})\cdot \hat{\bm B}({\bm r})$, where $\hat{{\bm m}}({\bm r})$ is the second quantized magnetic moment density of the BEC; $\hat{{\bm m}}({\bm r})=\sum_{\sigma,\sigma'}\hat{\Psi}^{\dagger}_{\sigma}({\bm r})[\bm \mu]_{\sigma,\sigma'}\hat{\Psi}_{\sigma'}({\bm r})$.   In the qubit basis and neglecting Zeeman energy shifts, which simply renormalize the energies of the flux qubit $\varepsilon\to \varepsilon'$,  the coupling takes the form of that given in Eq.\,\eqref{qubit-qubit Hamiltonian}, with 
\begin{equation}
\Omega=\sqrt{N}{\bm g}^{}_{\uparrow,\downarrow}\cdot{\bm \mu}_{\uparrow,\downarrow}, \quad {\bm g}^{}_{\sigma,\sigma'}=\int  d{\bm r}\, \phi^{*}_{\sigma}({\bm r}) {\bm B}({\bm r})\phi^{}_{\sigma'}({\bm r}); \label{gdef}
\end{equation}
here $\phi^{}_{\sigma}({\bm r})$ is the center of mass atomic wave function for the trapped atoms. For a SQUID of radius  $1\,\mu$m carrying a $1\,$mA current, and a center-to-center  BEC-SQUID separation of $50\, \mu$m,  $|{\bm g}^{}_{\uparrow,\downarrow}\cdot{\bm \mu}_{\uparrow,\downarrow}|\sim 100\,$Hz.  Therefore,  given a typical BEC has $N=10^{6}$ atoms, $T_{\rm R/2}$ is on the order of $10\,\mu$s, while current SQUIDs have decoherence times of the order of $1\,\mu$s \cite{KakuyanagiPRL2007}.  

Thus, from the results presented in the above, for the BEC to be a viable element of  qRAM, the Rabi coupling, the SQUID decoherence {times}, or some combination thereof would have to be increased by a factor of approximately 100, 
according to the above parameter estimates.  The simplest way to achieve this would be to move the BEC closer to the SQUID.  For instance, at a  distance of $10\, \mu$m the Rabi coupling would  increase by almost  two orders of magnitude \cite{prox}. 
There are, however, potentially detrimental effects caused by  having the BEC in such close proximity 
to a (comparatively hot) surface, such as increased heating or spin flips caused by thermal emission from the SQUID \cite{Cano,Kasch}. 

In summary, we find that the   
quantum memory in a hybrid system 
can be functional even if the time to transfer the state is of the same order as the smallest decoherence time. 
{Ideally coherence times or couplings would be large enough to perform a large number of qubit operations. But as quantum computing remains in its infancy, identifying  the applicable limits of  current technology to manipulate quantum information in hybrid systems is of major importance.}  
We further expect our results to remain generally true, with only minor quantitative  differences, for systems with different qubit-qubit or qubit-bath couplings than those considered here (for example dissipation into a sub-ohmic bath), because the ratio of transfer to decoherence times should be relatively insensitive to the details of the model.  A detailed confirmation of this is left for future work.  As  an interesting and currently feasible application, a hybrid BEC-SQUID qRAM system was introduced. We have demonstrated that with the present coherence times of SQUIDs, the derived effects of decoherence on the transfer fidelity need to be taken into account, to assess the functionality of the BEC-SQUID qRAM as a quantum memory device. 

\acknowledgments
This research was supported by the NRF of Korea, grant Nos. 2010-0013103 and 2011-0029541.


\begin{thebibliography}{0}
\bibitem{windowfunction} Explicitly, $W(t)=\Theta(\kappa,t)\Theta(\kappa,\tau-t)+\Delta^{}_{\rm or}\Theta(\kappa,-t)+\omega^{}_{\rm or}\Theta(\kappa,t-\tau)$, 
where $\Theta(\kappa,t)=\frac{1}{2}+\frac{1}{2}\tanh(\kappa t)$, $\tau$ is the time the two systems are on resonance (typically for half a Rabi cycle), and  $\omega^{}_{\rm or}$ is the off resonance (or) value in units of $\omega_{\rm m}$ (we set $\omega^{}_{\rm or}=\frac{1}{2}$ for all our calculations).  
The function $\Theta(\kappa,t)$ is a smoothed step function, such that $\lim_{\kappa\rightarrow\infty}\Theta(\kappa,t)=\Theta(t)$; different ramp times are obtained by varying  $\kappa$. For  simulations,  $\kappa=\omega_{\rm m}/8$. 
\bibitem{tempnote} {We have verified that, within numerical accuracy, the data points shown in the bottom parts of Figs.\,1 and 2 are left unchanged when the temperature is increased from $\beta^{-1}=\omega_{\rm m}/100$ to $\beta^{-1}=\omega_{\rm m}/10$.}
\bibitem{T1andT2times} In the literature one can find several definitions of decoherence times, which can differ slightly from the ones used here, e.g., by factors of two.
\bibitem{operatornote} For clarity, we note that the operator notation for SQUID magnetic field variables  exclusively refers to 
the flux qubit space, and there is no electromagnetic field quantization implied.
\bibitem{prox} {Note that 
the inhomogeneity of the magnetic field acting on the atoms when the SQUID is close to the BEC
does not affect the $\sqrt{N}$ enhancement of the dipole coupling, which only requires that a condensate 
is present in the trap.
The inhomogeneity of both the condensate density and the magnetic field is fully incorporated in the coupling matrix ${\bm g}^{}_{\sigma,\sigma'}$, defined in Eq.\,\eqref{gdef}. The BEC can therefore be of much larger spatial extent than the SQUID, and thus also have a conventionally small diluteness parameter.}
\end{thebibliography}
\end{document}